\documentclass[11pt,a4paper]{article}

\usepackage[T1]{fontenc}
\usepackage[utf8]{inputenc}

\usepackage[a4paper,margin=1in]{geometry}
\usepackage{setspace}

\usepackage{amsmath}
\usepackage{amssymb}

\usepackage{graphicx}
\usepackage{float}

\usepackage[version=4]{mhchem}
\usepackage{chemformula}

\usepackage{siunitx}
\usepackage{upgreek}
\usepackage{textcomp}

\usepackage{authblk}

\usepackage{xcolor}
\usepackage{verbatim}
\usepackage{ragged2e}

\usepackage[numbers,sort&compress]{natbib}
\usepackage[hidelinks]{hyperref}

\title{High-Q, high-power spin Hall nano-oscillators with thick NiFe layers enabled by SiC substrates}

\author[1,2,3]{Maha Khademi}
\author[1,4,*]{Akash Kumar}
\author[1]{Martina Ahlberg}
\author[1,5]{Artem Litvinenko}
\author[1,6]{Xiaotian Zhao}
\author[2]{Saroj P. Dash}
\author[1,7,8,*]{Johan \AA{}kerman}

\affil[1]{Department of Physics,
University of Gothenburg,
412 96 Gothenburg, Sweden}

\affil[2]{Department of Microtechnology and Nanoscience,
Chalmers University of Technology,
412 96 Gothenburg, Sweden}

\affil[3]{NanOsc AB,
Kista, 164 40, Sweden}

\affil[4]{Centre for Interdisciplinary and Convergent Technologies, Indian Institute of Technology Kharagpur, Kharagpur-721304, India}

\affil[5]{Department of Physics,
Oakland University,
Rochester, MI 48309, USA}

\affil[6]{Shenyang National Laboratory for Materials Science,
Institute of Metal Research,
Shenyang 110016, China}

\affil[7]{Center for Science and Innovation in Spintronics,
Tohoku University,
Sendai 980-8577, Japan}

\affil[8]{Research Institute of Electrical Communication,
Tohoku University,
Sendai 980-8577, Japan}

\affil[]{Correspondence:
\href{mailto:akash.kumar@physics.gu.se}
{akash.kumar@physics.gu.se};
\href{mailto:johan.akerman@physics.gu.se}
{johan.akerman@physics.gu.se}}

\date{}

\begin{document}

\justifying



\maketitle

\begin{abstract}

Spin Hall nano-oscillators (SHNOs) are nanoscale microwave sources with potential applications in local spin excitation, neuromorphic computing, and oscillator-based Ising machines. Further improvements in output power and coherence are essential for these applications, and increasing the thickness of the ferromagnetic layer offers a promising route toward this goal. However, this comes at the cost of higher drive currents, which increase Joule heating and can degrade device performance. Here, we show that this bottleneck can be overcome by SiC substrates. By comparing otherwise identical W/NiFe nanoconstriction SHNOs on $\mathrm{Al}_2\mathrm{O}_3$, Si, and SiC, we find that the high thermal conductivity of SiC enables stable auto-oscillations in NiFe layers as thick as \SI{10}{nm}, while the optimum performance occurs at \SI{7}{nm}. The best SiC device reaches an average output power of \SI{13.1}{pW} and a mean quality factor of $Q = 23{,}000$, representing the highest reported simultaneous output power and coherence for a single free-running SHNO.  These results establish SiC and thermal substrate engineering as a route to thick-ferromagnet spintronic oscillators for coherent microwave generation in nanoscale devices.

\end{abstract}


%
%
%
%
%
%
%
%

\section{Introduction}

Spintronic nano-oscillators \cite{Myers1999,Katine2000,Liu2012prl,demidov2012nm,kumar2024mutual} provide a versatile platform for microwave signal generation~\cite{dieny2020natelectron} and have recently gained interest as hardware primitives for neuromorphic computing~\cite{grollier2020nt,torrejon2017neuromorphic}, oscillator-based information processing~\cite{csaba2020coupled,borders2019integer,manipatruni2018nt}, and Ising machines~\cite{houshang2022prappl,yogendra2016ieee}. Their strong intrinsic nonlinearity \cite{slavin2009nonlinear,dvornik2018origin}, compatibility with CMOS processing \cite{zahedinejad2018cmos}, and ability to mutually synchronize or injection-lock to external stimuli \cite{Demidov2014b,torrejon2017neuromorphic,romera2022binding,sharma2021electrically} make them attractive for scalable networks of coupled oscillators. Among the various implementations, heavy metal/ferromagnet (HM/FM) based nanoconstriction (NC) spin Hall nano-oscillators (SHNOs)~\cite{demidov2012nm,Duan2014,Demidov2014} have emerged as a leading architecture due to their straightforward nanofabrication \cite{kumar2022fabrication,behera2024ultra}, wide frequency tunability \cite{awad2020apl,durrenfeld2017nanoscale,behera2022energy}, robust mutual synchronization \cite{awad2017natphys,zahedinejad2020nt,kumar2024mutual,kumar2025spinwaves}, and electrical \cite{fulara2020nt,Zahedinejad2022natmat,kumar2022fabrication,choi2022voltage,Khademi2023Large} and optical \cite{muralidhar2022optothermal} controllability. In SHNOs, spin currents generated in the HM by the spin Hall effect and related spin–orbit phenomena \cite{Dyakonov1971pl,Hirsch1999,sinova2015revmodern} exert spin-orbit torques on the FM and sustain magnetization auto-oscillations, which are converted into microwave voltage signals through the anisotropic magnetoresistance (AMR) \cite{demidov2012nm,Demidov2014,Duan2014}. Optimal SHNO operation hence requires a balanced current flow between the HM providing spin-orbit torque and the FM giving the AMR~\cite{prenat2015ultra}.

A promising route toward higher-performance SHNOs is to increase the FM thickness. Early NC-SHNOs were primarily based on Pt/NiFe bilayers with NiFe thicknesses of \num{3}--\SI{5}{nm}~\cite{Demidov2014,durrenfeld2017nanoscale,awad2017natphys,zahedinejad2020nt,kumar2023robust,awad2020apl}, typically yielding output powers ($P$) 
in the \SI{}{pW} range and linewidths ($\Delta f$) exceeding \SI{1}{MHz} for single free-running devices~\cite{Demidov2014,awad2017natphys,durrenfeld2017nanoscale,awad2020apl}. 
Increasing the FM thickness offers several important advantages: \emph{i}) larger AMR, lower damping due to smaller interfacial contributions to the effective damping, and increased current flow through the FM, which enhance the coherence and the electrical microwave output power; \emph{ii}) a larger magnetic mode volume, which can reduce phase noise and improve coherence; \emph{iii}) a larger dynamic magnetic moment, which increases the emitted microwave electromagnetic field; and \emph{iv}) stronger dipolar coupling between neighboring oscillators, an essential ingredient for robust mutual synchronization. The stronger microwave near-field could be particularly useful for the local excitation and control of nearby nanoscale spin systems, including spin defects and qubit platforms \cite{kraus2014room,Koehl2011nature}. 
The thick-FM SHNO strategy also connects to recent spin-current-driven magnonics experiments in large magnetic volumes, where damping compensation and auto-oscillations were demonstrated in a \SI{150}{nm}-thick YIG film by self-localizing the excited magnons beneath a Pt spin-current injector~\cite{schlitz2025natcomm}.

The central challenge is that a larger magnetic volume must be damping compensated by the spin currents generated in the adjacent HM layer. In metallic NC-SHNOs, this scaling problem is compounded by current redistribution between the HM and FM layers, and Joule heating caused by the large current densities  required for auto-oscillation. 
The resulting temperature rise reduces the saturation magnetization~\cite{Demidov2014}, increases damping~\cite{Yin2017, Zhao2016}, can reduce or modify the spin Hall efficiency~\cite{Wong2020}, suppresses AMR~\cite{Berger1988} and the microwave readout signal, and enhances thermal fluctuations. 
At high drive currents, Joule heating and incoherent mode interactions can also trigger current-induced magnetic instabilities, including vortex nucleation~\cite{Petit-Watelot2012,machado2012}, mode hopping~\cite{Chen2019pra}, and chaotic magnetization dynamics~\cite{Chen2019prb}. These coupled effects broaden the linewidth through nonlinear amplitude--phase coupling~\cite{Tiberkevich2008prb,kim2008prl,slavin2009nonlinear} and limit access to the high-current regimes where high output power and low linewidth coexist~\cite{litvinenko2023phase,muralidhar2022optothermal}. 
Efficient thermal management is therefore a key design parameter for thick-FM SHNOs.

To date, SHNOs have mainly been fabricated on sapphire (Al$_2$O$_3$)~\cite{durrenfeld2017nanoscale,awad2017natphys} and high-resistivity silicon~\cite{awad2017natphys,durrenfeld2017nanoscale,zahedinejad2020nt,chen2020GMR,kumar2023robust,kumar2025spinwaves}. Sapphire combines excellent electrical insulation with very high thermal conductivity at cryogenic temperatures, which has made it attractive for low-temperature nanodevices~\cite{Kawabata2025prres}; however, at room temperature its thermal conductivity is only moderate. Silicon offers CMOS compatibility~\cite{zahedinejad2018cmos}, but often requires additional oxide seed layers to suppress current shunting because its resistivity is limited to about 10--\SI{40}{\kilo\ohm}$\cdot$cm in typical high-resistivity substrates~\cite{behera2024ultra}. Silicon carbide (SiC) is a compelling alternative for room-temperature, high-current SHNO operation because it combines high thermal conductivity, high electrical resistivity, and compatibility with Si-based microelectronics~\cite{SiChighpower,liu2015silicon}. SiC is also an emerging quantum-photonic and spin-defect platform, hosting optically addressable color centers such as silicon vacancies, divacancies, and nitrogen-vacancy centers whose spin states can be optically initialized and read out, and coherently manipulated using microwave fields~\cite{guidry2022quantum,castelletto2022silicon,rahmouni2024entangled,kraus2014room,Koehl2011nature,bejarano2024parametric}. This combination makes SiC attractive not only as a heat-spreading substrate for high-current SHNO operation, but also as a possible platform for integrating nanoscale microwave-field sources with spin-defect and quantum devices. A necessary first step toward this longer-term direction is to establish whether SiC can support high-performance spintronic nano-oscillators.

Here, we demonstrate W/NiFe NC-SHNOs fabricated on $\mathrm{Al}_2\mathrm{O}_3$, Si, and SiC substrates and systematically compare their performance as a function of substrate, constriction width, and NiFe thickness. We show that the superior thermal and electrical properties of SiC enable stable high-current operation of thick-NiFe SHNOs and, for $t_{\mathrm{NiFe}}=\SI{7}{nm}$, grant access to a high-current, high-power regime consistent with an interior and expanding auto-oscillation mode that is not fully reached on the conventional substrates. 
Moreover, SiC allows auto-oscillations to be sustained in NiFe layers as thick as \SI{10}{nm}, substantially extending the accessible FM-thickness range of NC-SHNOs. 


We identify \SI{7}{nm} NiFe as the optimum thickness for the present W/NiFe stack, and observe that wider NCs give better performance. At high currents, the best device reaches an average output power of \SI{13.1}{pW} and a mean quality factor of $Q = 23{,}000$. This is, to our knowledge, the highest combination of power and $Q$ reported for a single free-running SHNO. The results establish thermal substrate engineering as an effective pathway toward high-coherence, thick-FM SHNOs for next-generation microwave sources and unconventional computing applications.

\section{Results and Discussion}

\subsection{Thermal Substrate Engineering for High-Current SHNO Operation}

\begin{figure}[htb]
\centering
\includegraphics[width=16 cm]{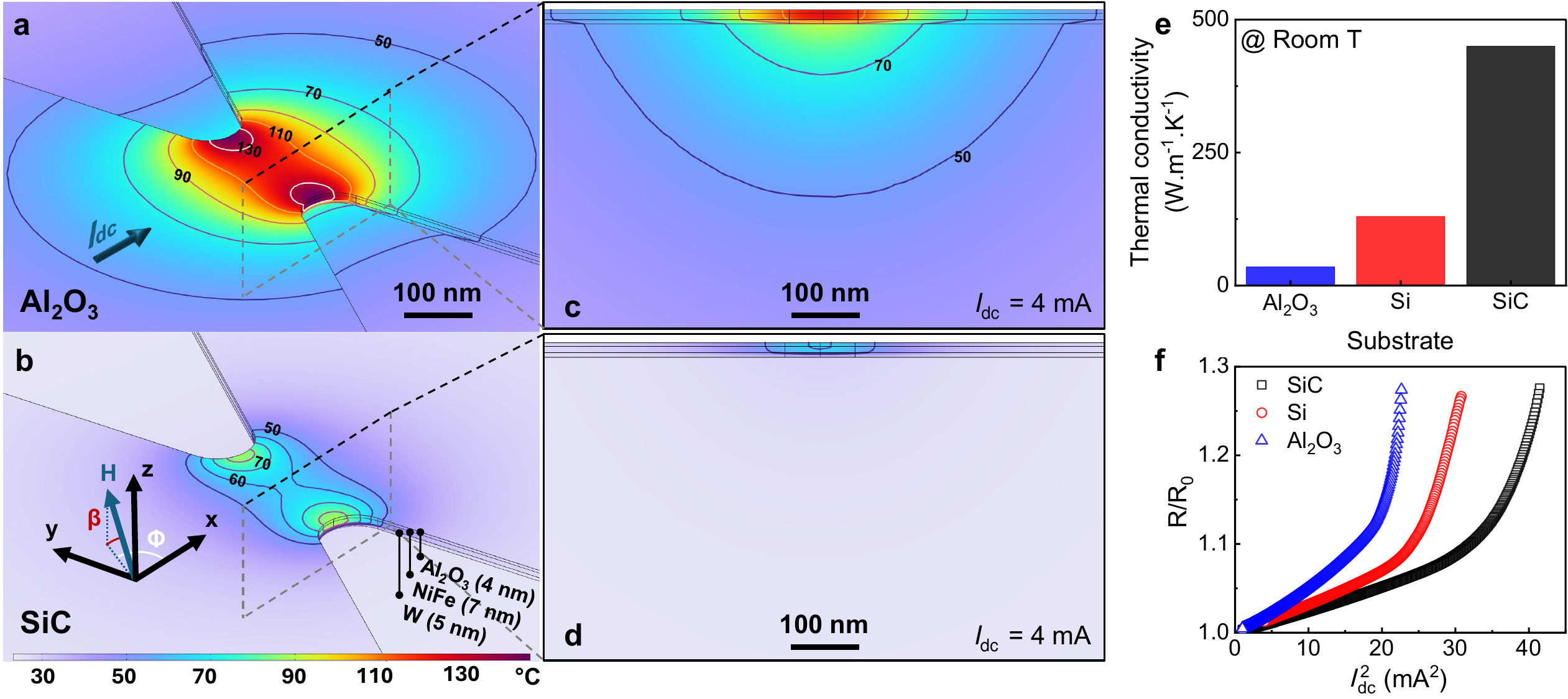}
\caption{Thermal properties and simulated heat spreading in \SI{200}{nm} SHNOs with $t_{\mathrm{NiFe}}=\SI{7}{nm}$.
(\textbf{a,b}) COMSOL temperature maps of SHNOs on (\textbf{a}) Al$_2$O$3$ and (\textbf{b}) SiC substrates at $I_{\mathrm{dc}}=\SI{4}{mA}$. The maps are plotted on the same temperature scale to enable direct comparison. The Al$_2$O$_3$ device develops a broad and strongly elevated temperature profile around the nanoconstriction, whereas the SiC device remains substantially cooler with a more spatially localized hot region. Isotherm contour lines indicate the local temperature distribution.
(\textbf{c,d}) Corresponding cross-sectional temperature maps for (\textbf{c}) Al$_2$O$_3$ and (\textbf{d}) SiC, illustrating heat spreading through the device stack and into the substrate.
(\textbf{e}) Room-temperature thermal conductivities of Al$_2$O$3$~\cite{Kawabata2025prres}, Si~\cite{thermalSi1,thermalSi2}, and SiC~\cite{thermalCSiC1,thermalCSiC2}.
(\textbf{f}) Normalized device resistance as a function of $I_{\mathrm{dc}}^{2}$, providing an experimental measure of current-induced self-heating and thermal runaway.
}
\label{fig:schematic}
\end{figure}

SHNOs were fabricated from W(\SI{5}{nm})/NiFe($t_{\text{NiFe}}$)/$\mathrm{Al}_2\mathrm{O}_3$(\SI{4}{nm}) trilayers, with $t_{\text{NiFe}} =$ 5, 7, 8, 9, and 10 nm. 
dc/rf magnetron sputtering was used to deposit the thin films on three substrates simultaneously: $c$-axis $\mathrm{Al}_2\mathrm{O}_3$, intrinsic (high-resistivity) Si ($\rho_{\text{Si}}>\SI{20}{\kilo\ohm} \cdot$cm), and high-purity 4H SiC. SHNO mesas with different constriction widths (20--\SI{200}{nm}) 
were patterned using e-beam lithography and etched by Ar-ion milling, and the widths were confirmed by scanning electron microscopy. The trilayers were also patterned into $4\times$\SI{14}{\micro\meter} microbars used for magnetoresistance and spin-torque ferromagnetic resonance (ST-FMR) measurements. More details on fabrication can be found in Ref.~\cite{kumar2022fabrication}. All measurements were performed at room temperature, and the power spectral density (PSD) maps were acquired in an applied field oriented along the out-of-plane angle $\beta=\SI{80}{\degree}$ and in-plane angle $\phi=\SI{19}{\degree}$. 

We first characterize the thermal environment of the nanoconstriction, since it sets the accessible current range for thick-FM SHNO operation. 
The thermal simulations were performed using COMSOL Multiphysics 6.3 for a W(\SI{5}{\nano\meter})/NiFe(\SI{7}{\nano\meter})/Al$_2$O$_3$(\SI{4}{\nano\meter}) nanoconstriction SHNO with a width of 200 nm, placed on a $50 \times 50 \times 50$ $\mu$m$^3$ substrate. The coupled Magnetic and Electric Fields (mef) and Heat Transfer in Solids (ht) modules were used to calculate the current distribution, Joule heating, and resulting temperature profiles. A fixed temperature of 293.15 K was imposed on all edges of the substrate, except the top surface containing the device. Constant thermal conductivities of  $\mathrm{Al}_{2}\mathrm{O}_{3}$, Si, and SiC substrates were set to $\kappa_{{\text{Al}_{2}\text{O}_{3}}}=$~\SI{35}{\watt}/m$\cdot$K~\cite{Kawabata2025prres}, $\kappa_{\text{Si}}=$~\SI{130}{\watt}/m$\cdot$K~\cite{thermalSi1,thermalSi2}, and $\kappa_{\text{SiC}}=$~\SI{450}{\watt}/m$\cdot$K~\cite{thermalCSiC1,thermalCSiC2}, respectively, as shown in Fig.~\ref{fig:schematic}(e). The following constant thermal conductivities were used for the W, NiFe, and $\text{Al}_{2}\text{O}_{3}$ layers: $\kappa_{\text{W}}=$~\SI{2}{\watt}/m$\cdot$K~\cite{han2017interface}, $\kappa_{\text{NiFe}}=$~\SI{17}{\watt}/m$\cdot$K~\cite{fangohr2011joule,ho1978thermal}, and $\kappa_{{\text{Al}_{2}\text{O}_{3}}}=$~\SI{2}{\watt}/m$\cdot$K~\cite{lee1995thermal}. Their electrical conductivities were: $\sigma_{\text{W}}=$~\SI{0.4606e6}{S/m}~\cite{mazraati2018auto,han2017interface}, $\sigma_{\text{NiFe}}=$~\SI{2.5e6}{S/m}~\cite{mazraati2018auto,fangohr2011joule}, and $\sigma_{{\text{Al}_{2}\text{O}_{3}}}=$~\SI{1e-12}{S/m}~\cite{li2000low,voigt2004electrical}.

Figure~\ref{fig:schematic}(a,b) shows the simulated temperature maps for \SI{200}{nm}-wide SHNOs with $t_{\text{NiFe}}=\SI{7}{nm}$ on Al$_2$O$_3$ and SiC substrates for an applied current ($I_{\text{dc}}$) of \SI{4}{mA}. 
The maps are plotted on the same temperature scale to enable direct comparison. In both devices, the hottest regions occur near the constriction edges, where the current density is largest. For Al$_2$O$_3$, the peak temperature reaches \SI{142}{\celsius}, and the isotherm contours indicate that much of the constriction interior remains above \SI{110}{\celsius}. In contrast, the SiC device remains substantially cooler, with most of the constriction region below \SI{60}{\celsius} and the temperature rise confined more closely to the current-crowding regions. The corresponding cross-sectional maps in Figure~\ref{fig:schematic}(c,d) further show that the substrate strongly modifies the heat-spreading geometry around the nanoconstriction, producing very different thermal landscapes for the high-current auto-oscillation modes. The different thermal profiles are consistent with the room-temperature thermal conductivities shown in Fig.~\ref{fig:schematic}(e). 
This hierarchy explains why Al$_2$O$_3$ provides the least favorable room-temperature thermal environment of the three substrates, whereas SiC enables substantially more efficient heat spreading and a larger accessible current range before thermal runaway.

We then experimentally examine the current-dependent resistance of \SI{200}{nm} NC-SHNOs on the three substrates as a probe of self-heating under dc operation. Figure~\ref{fig:schematic}(f) shows the normalized resistance plotted as a function of $I_{\mathrm{dc}}^{2}$. All devices initially show an approximately linear dependence on $I_{\mathrm{dc}}^{2}$, as expected for Joule heating, followed by a faster increase at higher current that signals the onset of thermal runaway. To limit possible heat-related device degradation, the current sweep was stopped once $R$ reached approximately 130\% of its initial value. The initial slope and the current at which the nonlinear increase begins differ strongly between substrates: the slope is largest for Al$_2$O$_3$, smallest for SiC, and intermediate for Si. The Al$_2$O$_3$ and Si devices therefore reach the onset of thermal runaway at substantially lower currents than the SiC device, in agreement with the simulated temperature profiles and the thermal-conductivity hierarchy in Figure~\ref{fig:schematic}(a--e). 

\subsection{Substrate-Dependent Auto-Oscillation Performance}


\begin{figure}[htb]
\centering
\includegraphics[width= 12 cm]{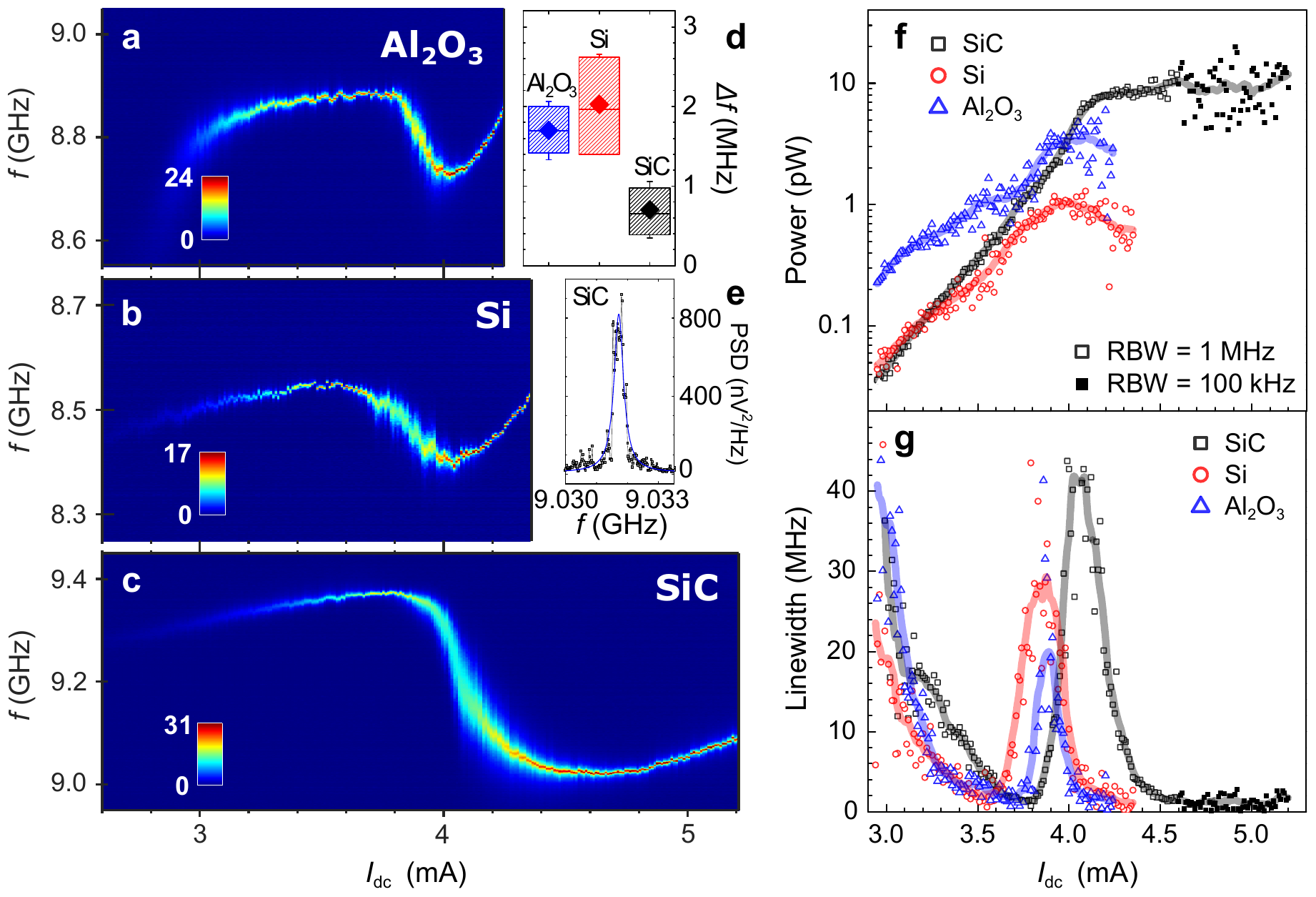}
\caption{Auto-oscillation characteristics of \SI{200}{nm} SHNOs with $t_{\text{NiFe}}=\SI{7}{nm}$ fabricated on different substrates. (\textbf{a--c}) PSD maps of devices on (\textbf{a}) Al$_2$O$_3$, (\textbf{b}) Si, and (\textbf{c}) SiC, measured in an external magnetic field of \SI{0.78}{T}. The numbers in the color bars indicate the upper limits of the plotted PSD scale in dB above the noise floor. (\textbf{d}) Linewidth statistics for Al$_2$O$_3$, Si, and SiC devices. The boxes span the 25th--75th percentiles, the center lines indicate the medians, the diamonds denote the means, and the whiskers represent the standard deviation. (\textbf{e}) Representative SiC spectrum and Lorentzian fit at $I{_\text{DC}}=\SI{4.69}{mA}$, giving a linewidth of $\Delta f=\SI{0.4}{MHz}$ and $Q = 24{,}000$. (\textbf{f,g}) Integrated output power (\textbf{f}) and linewidth (\textbf{g}) as functions of $I_{\text{dc}}$. Open and filled symbols denote measurements performed with resolution bandwidths (RBW) of \SI{1}{MHz} and \SI{100}{kHz}, respectively.}
\label{fig:200nm compare} 
\end{figure}

Next, we compare the substrate-dependent auto-oscillation dynamics of \SI{200}{nm} SHNOs with $t_{\mathrm{NiFe}}=\SI{7}{nm}$. Figure~\ref{fig:200nm compare}(a--c) shows current-dependent PSD maps for devices on Al$_2$O$_3$, Si, and SiC. The absolute auto-oscillation frequencies differ between the nominally identical stacks, consistent with modest substrate-dependent variations in the magnetic properties of the W/NiFe trilayers. ST-FMR measurements give effective magnetizations of $\mu_{0}M_{\mathrm{eff}}=0.835$, 0.800, and \SI{0.845}{T} for Al$_2$O$_3$, Si, and SiC, respectively; the lower $\mu_{0}M_{\mathrm{eff}}$ of the Si device is consistent with its lower auto-oscillation frequency range. Figure~\ref{fig:200nm compare}(d,e) summarizes the linewidth statistics and shows a representative Lorentzian fit to a high-coherence SiC spectrum, while Figure~\ref{fig:200nm compare}(f,g) compares the integrated output power and linewidth as functions of $I_{\mathrm{dc}}$. The resolution bandwidth (RBW) was reduced from \SI{1}{MHz} to \SI{100}{kHz} at the highest currents to resolve the narrowest linewidths of the SiC device.

The frequency evolution in Figure~\ref{fig:200nm compare}(a--c) is strongly nonmonotonic, reflecting the current-driven spatial evolution of the auto-oscillation mode in a tilted out-of-plane field~\cite{dvornik2018origin,awad2017natphys}. Close to threshold, the auto-oscillation nucleates from a localized linear mode near the constriction edge, where the demagnetizing field creates a spin-wave well and the current density is largest. As $I_{\mathrm{dc}}$ increases, the Oersted field and current-induced torques modify the effective mode landscape, and the pronounced red-shift and frequency minimum are consistent with the active mode moving away from the edge and toward the constriction interior. At still higher currents, the mode can expand beyond the immediate constriction region into adjoining parts of the device, increasing its precession volume; accessing this expanded-mode regime is the central reason that improved heat dissipation is beneficial for thick-FM SHNOs. This high-current regime is where the largest integrated power and the narrowest linewidth are obtained.

This mode evolution makes the substrate-dependent thermal landscape particularly important, since Joule heating becomes substantial in the same high-current range. The high-current slopes $\mathrm{d}f/\mathrm{d}I_{\mathrm{dc}}$ are approximately \SI{560}{\mega\hertz\per\milli\ampere}, \SI{400}{\mega\hertz\per\milli\ampere}, and \SI{150}{\mega\hertz\per\milli\ampere} for Al$_2$O$_3$, Si, and SiC, respectively. Although these slopes also depend on the Oersted field, nonlinear frequency shift, and detailed mode profile, their systematic substrate dependence indicates a strong thermal contribution. Self-heating changes the magnetic parameters, including $\mu_{0}M_{\mathrm{eff}}$, and thereby contributes to the current-induced frequency shift, consistent with earlier optothermal studies of NC-SHNOs~\cite{muralidhar2022optothermal}. The much smaller slope on SiC is therefore consistent with the weaker thermal feedback inferred from the $R(I_{\mathrm{dc}})$ data in Figure~\ref{fig:schematic}(f) and the simulated temperature maps in Figure~\ref{fig:schematic}(a--d).

For quantitative comparison, the best auto-oscillation characteristics were determined from spectral peaks with narrow linewidths and well-defined Lorentzian line shapes, as evaluated by the fit $R^{2}$-value. Specifically, the integrated output power $P$ and linewidth $\Delta f$ were obtained by averaging over the 25 spectral peaks with the narrowest linewidths that also fulfilled the Lorentzian fit-quality criterion. The quality factor $Q$ was averaged separately over the top 10 peaks within this subset, since $Q$ is more sensitive to linewidth outliers. The resulting linewidth statistics are shown in Figure~\ref{fig:200nm compare}(d). 
Figure~\ref{fig:200nm compare}(e) shows a representative SiC peak at $I_{\mathrm{dc}}=\SI{4.69}{mA}$, for which a Lorentzian fit gives $\Delta f=\SI{0.4}{MHz}$ and $Q = 24{,}000$. Individual selected spectra reach linewidths as low as 200 kHz and Q-factors up to 45,000.

The less favorable thermal environment of the Al$_2$O$_3$ and Si devices limits the available current range to below approximately \SI{4.3}{mA}, preventing these devices from fully reaching the high-current regime where maximum power and minimum linewidth coexist. Their average output powers are $P=\SI{3.8}{pW}$ and \SI{1.0}{pW}, respectively, with linewidths consistently larger than \SI{1}{MHz} and quality factors of about 6,500 and 6,300. 
In contrast, the efficient heat dissipation provided by SiC allows the device to reach the high-current expanding-mode regime while reducing thermally induced linewidth broadening~\cite{kim2008prl,slavin2009nonlinear}. Consequently, the SiC device achieves an average integrated output power of \SI{13.1}{pW}, a sub-MHz linewidth, and a mean quality factor of $Q=23{,}000$.
This represents, to our knowledge, the highest power--$Q$ combination reported for a single free-running SHNO.
Previous single free-running SHNOs have demonstrated output powers of approximately \SI{1}{pW} to \SI{9}{pW}, with quality factors ranging from about 300 to 1,000~\cite{Duan2014,fulara2019spin,chen2020GMR}. Although GMR-based SHNOs have achieved substantially higher output powers of approximately \SI{1.2}{nW}, their quality factors remain much lower, around 33~\cite{chen2020GMR}. Mutual synchronization has enabled higher quality factors, reaching approximately 170,000 in an $8\times8$ SHNO array and 79,000 in smaller synchronized networks~\cite{zahedinejad2020nt,kumar2023robust}. These comparisons highlight the ability of the present SiC-based SHNO to combine high output power and spectral coherence in a single free-running device, without relying on mutual synchronization.


Having established SiC as the superior substrate for high-current operation, we briefly 
comment on the difference between the Al$_2$O$_3$ and Si devices. Although one might expect Si to outperform Al$_2$O$_3$ because of its higher room-temperature thermal conductivity, the Al$_2$O$_3$ device produces larger $P$ and lower $\Delta f$. Simple dc current shunting into the Si substrate is unlikely to be the dominant explanation: the measured zero-current resistances are comparable, with $R_{0}=\SI{228}{\ohm}$ for Al$_2$O$_3$, \SI{258}{\ohm} for Si, and \SI{235}{\ohm} for SiC. The Si device also shows a smaller $d(R/R_{0})/dI_{\mathrm{dc}}^{2}$ than Al$_2$O$_3$ over the measured current range, indicating better thermal stability. The lower output power on Si is therefore more likely related to substrate-dependent magnetic and/or crystalline differences in the W/NiFe stack. This interpretation is consistent with the Si sample having the lowest $\mu_{0}M_{\mathrm{eff}}$ and the lowest auto-oscillation frequencies, which can shift the current range where the edge-localized mode evolves into the high-power interior/expanding regime. 

\begin{figure}[H]
    \centering
    \includegraphics[width= 5.4 cm]{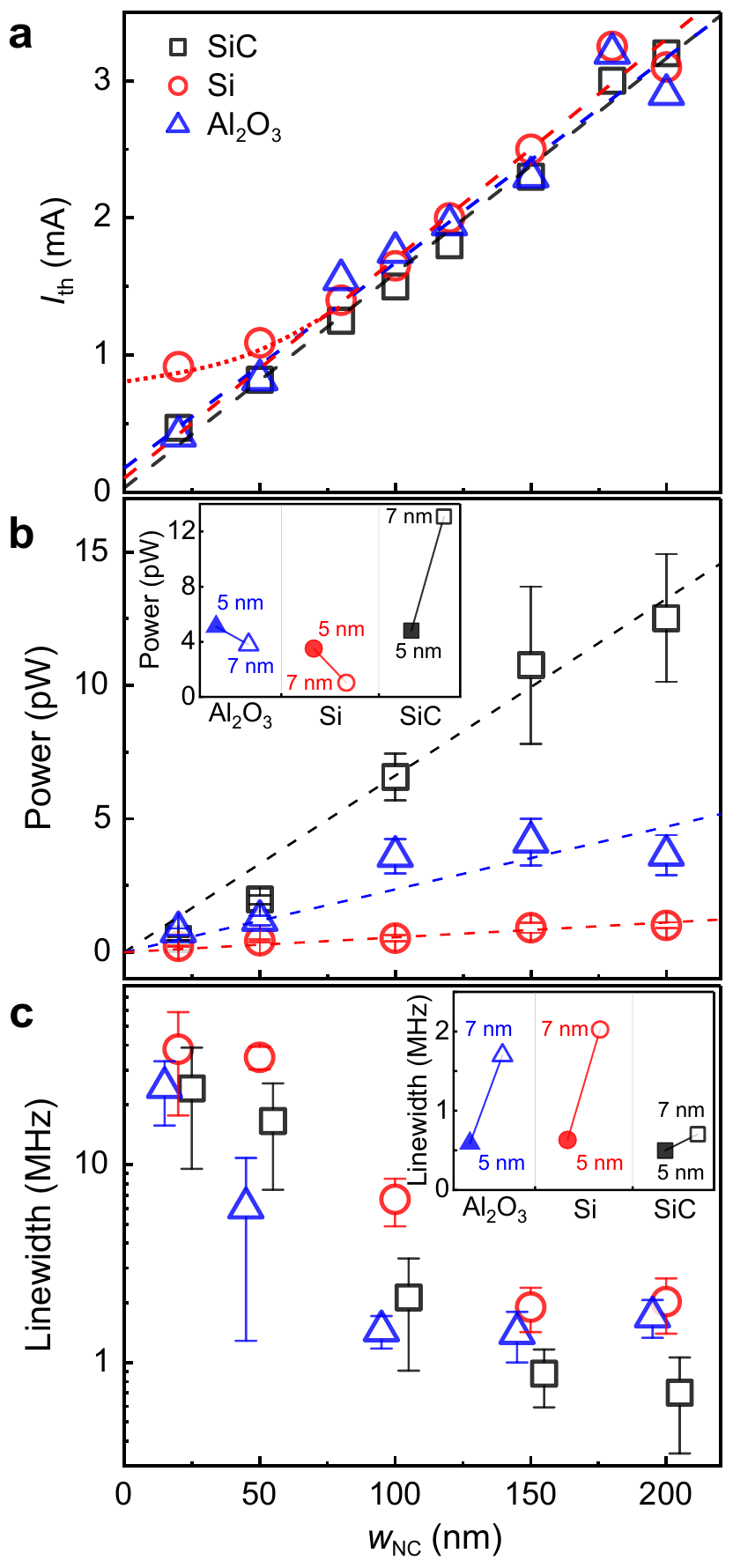}
    \caption{Nanoconstriction-width dependence of $t_{\text{NiFe}}=\SI{7}{nm}$ SHNOs on different substrates. (\textbf{a}) Threshold current ($I_{\text{th}}$), (\textbf{b}) integrated output power, and (\textbf{c}) linewidth as functions of nanoconstriction width ($w_{\text{NC}}$) for SHNOs fabricated on SiC, Si, and Al$_2$O$_3$ substrates. Error bars represent the standard deviation of the averaged data points. The dashed lines in (\textbf{a,b}) are guides to the eye. Data in (\textbf{c}) are horizontally offset for clarity.  The inset in (\textbf{b}) compares the integrated output power of \SI{200}{nm} SHNOs with $t_{\text{NiFe}}=\SI{5}{nm}$ and \SI{7}{nm}, while the inset in (\textbf{c}) shows the corresponding linewidth comparison. The measurements were performed in external magnetic fields of \SI{0.77}{T} for $t_{\text{NiFe}}=\SI{5}{nm}$ and \SI{0.78}{T} for $t_{\text{NiFe}}=\SI{7}{nm}$.}
    \label{fig:NC dependent}
\end{figure}


We now turn to the NC-width ($w_{\text{NC}}$) dependence of the device performance. The threshold current ($I_{\text{th}}$), power, and linewidth of SHNOs with $w_{\text{NC}}=$ 20--\SI{200}{nm} are compared in Fig.~\ref{fig:NC dependent}(a--c). 
As expected, the threshold current scales linearly with NC width for all substrates (Fig.~\ref{fig:NC dependent}(a)), except for 20 and \SI{50}{nm} devices on Si substrate where excessive current shunting prevents further improvement~\cite{behera2024ultra, awad2020apl}. The observation that $I_{\text{th}}$ of wide NC devices on Si follow the same trend as for the other substrates supports our conclusion above, that current shunting does not dominate the behavior of \SI{200}{nm} SHNOs. 
The power (Fig.~\ref{fig:NC dependent}(b)) improves with $w_{\text{NC}}$ for all devices, while the linewidth (Fig.~\ref{fig:NC dependent}(c)) decreases and reaches a substrate-dependent minimum. It is clear that SiC outperforms the two conventional substrates in terms of power for $w_{\text{NC}} \geq$~\SI{50}{nm}, while the most evident linewidth benefit appears for the \SI{150}{\nano\meter} and \SI{200}{\nano\meter} constrictions. 
The larger contact area between wider NCs and the substrate utilizes SiC’s high thermal conductivity more effectively, which enables more favorable operation conditions.

We also compare the performance of \SI{200}{nm} SHNOs with two different NiFe thicknesses, $t_{\text{NiFe}} = 5$ and \SI{7}{nm} in the insets in Fig.~\ref{fig:NC dependent}(b,c). The thinner NiFe devices are almost independent of substrate. 
Nonetheless, for SiC a clear improvement in power is achieved by increasing $t_{\text{NiFe}}$, while the performance of $\text{Al}_2\text{O}_3$ and Si devices worsens. In terms of the linewidth, both the $\text{Al}_2\text{O}_3$ and Si substrates exhibit a significant broadening when the NiFe thickness is increased from 5 nm to 7 nm. In contrast, the linewidth of devices fabricated on SiC remains nearly unchanged over the same thickness range.
The basis of optimal SHNO operation is a well-balanced current distribution between the HM and FM layers (and a perfectly insulating substrate). With increasing NiFe thickness, more current flows through the FM and less in the HM, reducing the spin-orbit torques and increasing the threshold current, thus moving the optimal expanded mode region to higher $I_{\mathrm{dc}}$. Sufficient $I_{\mathrm{dc}}$ can only be reached if the substrate provides effective heat transfer. Thus, only SiC-based devices can fully leverage both the larger magnetic volume and the AMR. In addition, the SiC $t_{\text{NiFe}}=\SI{7}{nm}$ SHNO offers further advantage through its enhanced current-induced frequency tunability. Specifically, it exhibits a tunability of approximately \SI{400}{MHz}, significantly outperforming the $t_{\text{NiFe}}=\SI{5}{nm}$ devices, which achieve only about \SI{200}{MHz}. 

\subsection{Optimization of NiFe Thickness on SiC}

\begin{figure}[htb]
    \centering
    \includegraphics[width= 16 cm]{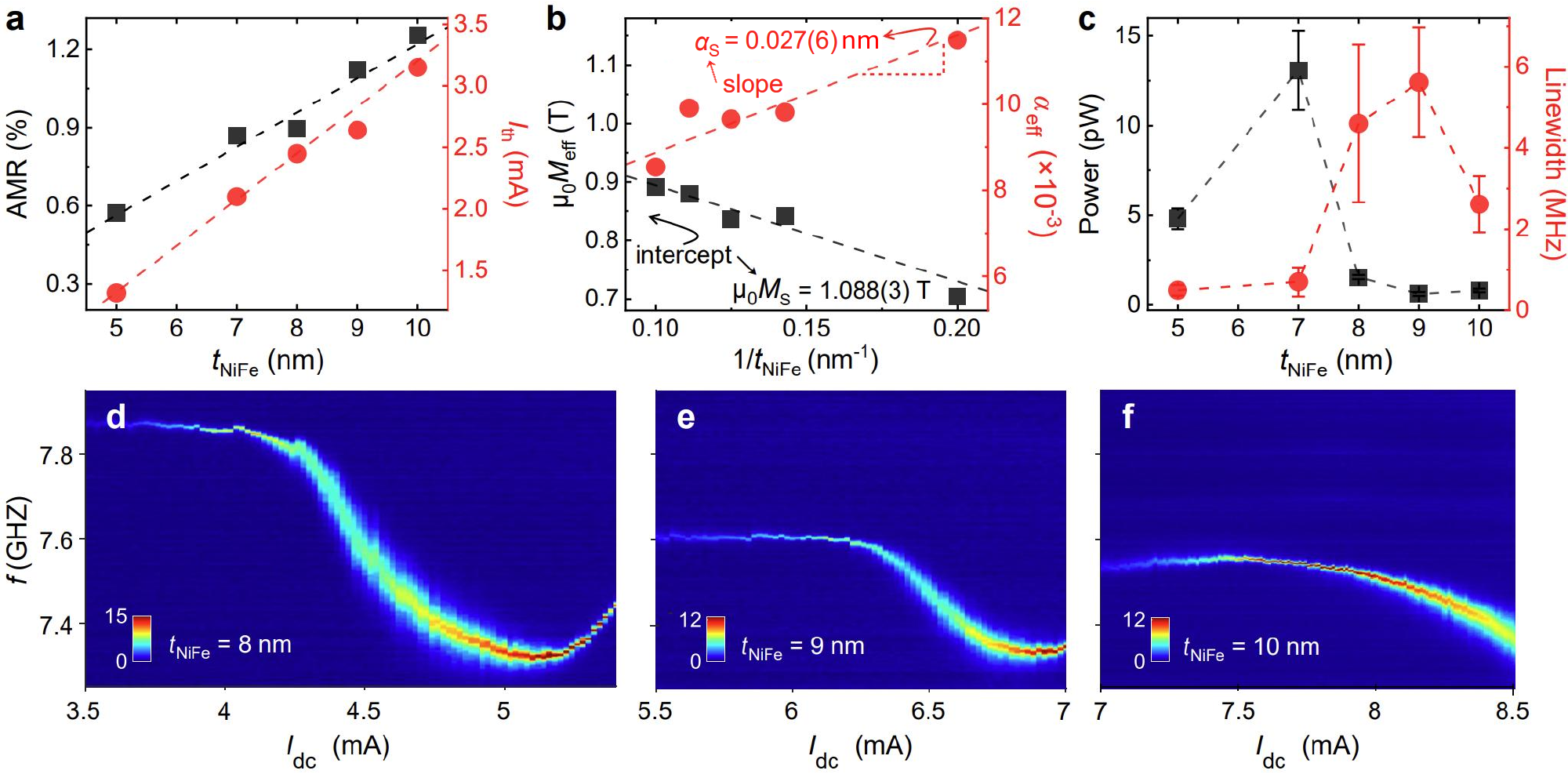}
    \caption{NiFe-thickness dependence of magnetic properties and auto-oscillation performance in \SI{200}{nm} SHNOs on SiC substrates. (\textbf{a}) AMR ratio (black squares) and threshold current (red circles) as functions of NiFe thickness. (\textbf{b}) Effective magnetization ($\upmu_0 M_{\mathrm{eff}}$) and Gilbert damping ($\alpha_{\text{eff}}$) extracted from ST-FMR measurements. The slope of the damping gives the surface contribution, $\alpha_{\text{S}}$ = \SI{0.027}{nm}. The extrapolated bulk saturation magnetization is $\upmu_{0}M_{\text{S}} = \SI{1.088}{T}$. (\textbf{c}) Integrated output power and linewidth as functions of NiFe thickness, showing optimum performance at $t_{\mathrm{NiFe}}=\SI{7}{nm}$. (\textbf{d--f}) PSD maps of \SI{200}{nm} SHNOs with $t_{\mathrm{NiFe}}=\SI{8}{nm}$, \SI{9}{nm}, and \SI{10}{nm}, respectively, demonstrating auto-oscillations in thicker NiFe layers on SiC. The maps were measured at applied fields of \SI{0.6}{T} (\SI{8}{nm} and \SI{9}{nm}), and \SI{0.56}{T} (\SI{10}{nm}).}
    \label{fig:thicker NiFe}
\end{figure}

Up to this point, we have shown that wide NC-SHNOs with $t_{\mathrm{NiFe}}=\SI{7}{nm}$ on SiC substrates exhibit stable auto-oscillation, high output power, and high quality factor. We now investigate how the magnetic properties and auto-oscillation dynamics evolve with NiFe thickness, focusing on SiC-based devices with $t_{\mathrm{NiFe}}\leq\SI{10}{nm}$. As the strength of the electrical auto-oscillation signal should be linear in AMR, Figure~\ref{fig:thicker NiFe}(a) shows the AMR ratio, determined from the angular dependence of $R$ in an applied field of \SI{0.2}{T}. 
The AMR increases approximately linearly with NiFe thickness, since a larger fraction of the total current flows through the thicker FM layer and contributes to the magnetoresistive readout. Thus, the maximum value of \SI{1.25}{\percent} is obtained for the largest $t_{\mathrm{NiFe}}$. The corresponding reduction of current in the W layer reduces the spin-orbit torque efficiency and raises the threshold current, which is also shown in Figure~\ref{fig:thicker NiFe}(a). Doubling $t_{\mathrm{NiFe}}$ increases $I_{\mathrm{th}}$ by a factor of approximately 2.5. 

Figure~\ref{fig:thicker NiFe}(b) presents the effective magnetization and Gilbert damping ($\alpha_{\text{eff}}$) as functions of inverse NiFe thickness. Both quantities were extracted from ST-FMR measurements and they are approximately linear in $1/t_{\text{NiFe}}$. Extrapolation to $1/t_{\mathrm{NiFe}}\rightarrow0$ gives a bulk saturation magnetization of $\upmu_{0}M_{\mathrm{s}}=\SI{1.088}{T}$, while the slope of $\alpha(1/t_{\mathrm{NiFe}})$ gives a surface-damping parameter of $\alpha_{\mathrm{s}}=\SI{0.027}{\nano\meter}$~\cite{barati2014gilbert,tserkovnyak2005nonlocal}. Following Eq.~(1) of Ref.~\cite{behera2022energy}, the effective spin Hall torque efficiency is estimated to lie in the range $-0.4$ to $-0.6$, using literature resistivity values of $\rho_{\mathrm{W}}=\SI{300}{\micro\ohm\centi\meter}$~\cite{petroff1973microstructure} and $\rho_{\mathrm{NiFe}}=\SI{40}{\micro\ohm\centi\meter}$~\cite{nahrwold2010structural}.

Finally, we evaluate auto-oscillations in thicker NiFe layers. Figure~\ref{fig:thicker NiFe}(d--f) show representative PSD maps for \SI{200}{nm} SHNOs with $t_{\mathrm{NiFe}}=\SI{8}{nm}$, \SI{9}{nm}, and \SI{10}{nm}. 
The current-dependent mode evolution changes markedly with NiFe thickness, reflecting the different magnetic energy landscapes, threshold currents, and current partitioning between the W and NiFe layers. Under optimized field conditions, the strongest signals for $t_{\mathrm{NiFe}}=\SI{8}{nm}$, \SI{9}{nm}, and \SI{10}{nm} reach approximately 15, 19, and \SI{11}{dB} above the noise floor, respectively. This is substantially below the \SI{31}{dB} obtained for the \SI{7}{nm} device. Figure~\ref{fig:thicker NiFe}(c) further shows that the thicker devices suffer from reduced output power and broadened linewidth. Thus, while SiC enables auto-oscillations in NiFe layers as thick as \SI{10}{nm}, the highest performance is still obtained at $t_{\mathrm{NiFe}}=\SI{7}{nm}$. Thicker NiFe layers increase the AMR and magnetic volume, but they also divert more current from the W layer into the NiFe layer and raise $I_{\mathrm{th}}$, making it increasingly difficult to reach the expanded high-current regime associated with maximum output power and minimum linewidth. The optimum thickness is therefore set by a trade-off between increasing AMR/magnetic volume and maintaining sufficient current in the W layer for efficient spin-orbit torque. Further improvements for $t_{\mathrm{NiFe}}>\SI{7}{nm}$ will likely require additional current-distribution engineering, for example through the W thickness, resistivity, or interface design.

\section{Conclusions}

To summarize, we have demonstrated W(\SI{5}{nm})/NiFe($t_{\mathrm{NiFe}}$) 
nanoconstriction SHNOs with $t_{\mathrm{NiFe}}=\SIrange{5}{10}{nm}$ on Al$_2$O$_3$, Si, and SiC substrates. By combining thermal simulations, current-dependent resistance measurements, and auto-oscillation spectroscopy, we show that SiC provides a substantially improved thermal environment for high-current SHNO operation. This enhanced heat dissipation extends the accessible current range and enables stable auto-oscillations in NiFe layers as thick as \SI{10}{nm}. The best performance is obtained for $t_{\mathrm{NiFe}}=\SI{7}{nm}$ on SiC, where the \SI{200}{nm} device reaches an average integrated output power of \SI{13.1}{pW} and a mean quality factor of $Q= 23{,}000$. To our best knowledge, this is a record power--$Q$ combination for a single free-running SHNO. 


These results establish SiC as a functional substrate platform for high-power, high-coherence SHNOs with thicker ferromagnetic layers. The benefits should be particularly relevant for oscillator chains and arrays, where large total drive currents and efficient heat spreading are essential for robust mutual synchronization. More broadly, the combination of high thermal conductivity, high electrical resistivity, and compatibility with Si-based and spin-defect technologies makes SiC an attractive substrate for future spintronic microwave sources, oscillator networks, and nanoscale microwave-field devices.

\medskip

\medskip
\textbf{Acknowledgements} \par 
This work was supported by the Swedish Research Council (VR Grant No.~2024-01943), the Knut and Alice Wallenberg Foundation (Grant No.~2023.0285), and the European Union’s Horizon 2020 Research and Innovation Programme through the Marie Sk\l{}odowska-Curie under Grant 955671 “SPEAR”. This work was performed in part at Myfab Chalmers. XZ acknowledges funding from National Nature Science Foundation of China under project No. 52271193.

\medskip
\textbf{Author Contributions} \par 
X.Z., A.K., and J.\AA{}. initiated the idea. M.K. prepared the thin films and fabricated the devices. M.K. performed all electrical auto-oscillation, and spin-orbit torque measurements with input from A.K. The data analysis was done by M.K. with assistance of M.A. and inputs from A.K. M.K. performed the COMSOL simulations with input from A.L. 
M.K. and M.A. were responsible for drafting the manuscript, with all authors contributing to writing the paper. All authors contributed to the data analysis and interpretation of the results.

\medskip

%
\bibliographystyle{MSP}
\bibliography{references.bib}

\end{document}